%% file: paper.tex
\documentclass[acmsmall,10pt,screen]{acmart}

\usepackage{todonotes}
\usepackage{csquotes}
\usepackage{mathpartir}
\usepackage{simplebnf}
\usepackage{amssymb}
\usepackage{amsfonts}
\usepackage{amsmath}
\usepackage{hyperref}

\title{One Weird Trick to Untie Landin's Knot}

\input{setup}
\input{config}

\author{Paulette Koronkevich}
\orcid{0000-0003-0325-3305}
\affiliation{%
  \institution{University of British Columbia}
  \country{Canada, }
  \href{mailto:pletrec@cs.ubc.ca}{pletrec@cs.ubc.ca}}%
\email{pletrec@cs.ubc.ca}
\author{William J. Bowman}
\orcid{0000-0002-6402-4840}
\affiliation{%
  \institution{University of British Columbia}
  \country{Canada, }
  \href{mailto:wjb@williamjbowman.com}{wjb@williamjbowman.com}}%
\email{wjb@williamjbowman.com}
\authorsaddresses{}

\acmConference{HOPE'23}{September 4--9, 2023}{Seattle, Washington, USA}
\acmBooktitle{Workshop on Higher-order Programming with Effects}
\acmYear{2023}
\acmMonth{9}
\setcopyright{none}
\acmDOI{}
\acmISBN{}
\acmPrice{}

\begin{document}
\begin{abstract}
  \input{abstract.tex}
\end{abstract}

\maketitle

\input{intro2.tex}
\input{proposal.tex}

\bibliographystyle{ACM-Reference-Format}
\bibliography{all}

\appendix
\input{appendix.tex}

\end{document}

%% file: setup.tex
\setcopyright{none}

\usepackage{multicol}
\usepackage{xspace}
\usepackage{enumitem}
\usepackage{tikz}
\usetikzlibrary{tikzmark}
\usepackage{pdflscape}
\usepackage{hyperref}
\usepackage{doi}

%% file: config.tex
\usepackage{simplebnf}

\NewCommandCopy{\oldbnfgrammar}{\bnfgrammar}
\NewCommandCopy{\endoldbnfgrammar}{\endbnfgrammar}

\RenewDocumentEnvironment{bnfgrammar}{O{lrclr} O{\|\|} O{\|}}
{\begin{oldbnfgrammar}[#1][#2][#3]}
{\end{oldbnfgrammar}}

\usepackage{semantic}
\mathlig{!->}{\mapsto}
\mathlig{|->}{\mapsto}
\mathlig{<-|}{\mapsfrom}
\mathlig{|->*}{\mapsto^{*}}
\mathlig{|[}{\llbracket}
\mathlig{|]}{\rrbracket}
\mathlig{|>}{\vartriangleright}
\mathlig{|>*}{\vartriangleright^{*}}
\mathlig{<|}{\vartriangleleft}
\mathlig{=ciu}{\overset{ciu}{\approx}}
\mathlig{=ctx}{\overset{ctx}{\approx}}
\mathlig{=?=}{\overset{?}{=}}
\mathlig{=[}{\sqsubseteq}

\newcommand{\mmathbf}[1]{\ensuremath{\mathbf{#1}}}
\newcommand{\subst}[3]{{#1}[{#2}/{#3}]}

\reservestyle{\asmc}{\mmathbf}
\asmc{pack[pack],
  proj1[proj_{1}\;],
  proj2[proj_{2}\;],
  fg[full\text{-}ground\;],
  deref[deref\;],
  new[new\;],
  if[if\;],
  then[\;then\;],
  else[\;else\;],
  let[let\;],
  in[\;in\;],
  kwthen[then],
  kwelse[else],
  begin[begin\;],
  lam[\boldsymbol{\lambda}],
  eq?,neq,lt,lte,gt,gte,
  Ref[Ref\;],Type[Type],Lit,Label,
  rsp,rax,r8,r9,r10,rbp,
  inf,-inf,
  right[right\;],left[left\;],max,
  Int,U[U\;],Ptr[Ptr\;],*,Top,empty[\epsilon],
  forall[\boldsymbol{\forall}\;],
  exists[\boldsymbol{\exists}\;],
  F[F\;],
  kwF[F],
  kwU[U],
  kwPtr[Ptr],
  ?[\boldsymbol{?}],
  view[view\;],
  kwview[kwview],
  as[\;as\;],
  ass[as\;],  
  mu[\boldsymbol{\mu}],
  Reft[Ref_{type}\;], Clost[Clos_{type}\;],    
  Null,
  Update[Update\;],
  Focus[Focus\;]
  Zero?[zero?],
  Positive?[positive?],
  Negative?[negative?],
  Time[time],
  +[\;\boldsymbol{+}\;],
  *[\;\boldsymbol{*}\;],
  -[\;\boldsymbol{-}\;],
  <=[\;\boldsymbol{\leq}\;],
  kw+[\boldsymbol{+}],
  kw*[\boldsymbol{*}],
  kw-[\boldsymbol{-}],
  kw<=[\boldsymbol{\leq}],
  hole[\cdot],
  bind[bind\;],
  kwbind[bind],
  kwreturn[return],
  returnl[return_{\mathit{l}}\;],
  returni[return_{\mathit{i}}\;],
  returns[return_{\mathit{s}}\;],
  Nat,
  Clk,
  Bool,
  Tl[T_{\mathit{l}}\;],
  Tl'[T_{\mathit{l'}}\;],
  Ts[T_{\mathit{s}}\;],
  Ti[T_{\mathit{i}}\;],
  begin[begin\;],
}

%% file: abstract.tex
In this work, we explore Landin's Knot, which is understood as a pattern for
encoding general recursion, including non-termination, that is possible after
adding higher-order references to an otherwise terminating language.
We observe that this isn't always true---higher-order references, by themselves,
don't lead to non-termination.
The key insight is that Landin's Knot relies not primarily on references storing
functions, but on unrestricted quantification over a function's environment.
We show this through a closure converted language, in which the function's
environment is made explicit and hides the type of the environment through
impredicative quantification.
Once references are added, this impredicative quantification can be exploited to
encode recursion.
We conjecture that by restricting the quantification over the environment,
higher-order references can be safely added to terminating languages, without
resorting to more complex type systems such as linearity, and without
restricting references from storing functions.

%% file: intro2.tex
\section{Introduction}
How do we add \textit{higher-order references} to an otherwise pure functional
language?
The problem is that higher-order references (that is, mutable references that can store
higher-order functions) lead to Landin's Knot, which breaks strong normalization.
While many languages value the expressivity of higher-order references,
some languages, such as dependently-typed languages, value strong normalization
as well.
Existing languages often recover strong normalization by using
\textit{linear types}, which carefully track the usage of
references in the type system \cite{ahmed2007,krishnaswami2015}.
However, linear types are typically difficult to integrate into existing type systems,
in particular, dependent types \cite{mcbrideplenty}.
Quantitative Type Theory (QTT) \cite{atkey2018} is one language that succeeds at integrating
linear and dependent types, but it aims to provide a framework for reasoning
about resources in general, which seems a heavyweight solution to ruling out non-termination.

To explore alternative methods for adding higher-order references to pure functional
languages, let's study Landin's Knot in detail.
Landin's Knot \cite{landin1964} refers to encoding recursion through backpatching---updating
a mutable reference to create a cyclic data structure.
In this case, the cyclic data structure happens to be a closure, and a reference
is updated to contain the closure itself, enabling recursion.
This idea is illustrated through the following program in the simply typed
lambda calculus (STLC) with references, which diverges.

\vspace{-3ex}

\begin{displaymath}
  \begin{array}[t]{l}
    id :  \<Nat> \to \<Nat> \\
    id = (\<lam> x . x) \\
    r : \<Ref> (\<Nat> \to \<Nat>) \\
    r = \<new> id \\
    f : \<Nat> \to \<Nat> \\
    f = (\<lam> x . ((\<deref> r) \ x))\\
    r := f; f \ 0
  \end{array}
\end{displaymath}

\vspace{-1ex}

\noindent The function $f$ closes over the reference $r$, and calls whatever
function is stored in $r$.
Initially, $r$ contains an arbitrary function of the right type, but is later
updated to contain $f$ itself.
After the update, $f$ diverges when called.

The literature often attributes Landin's Knot merely to be due to references
\emph{storing} functions.
\begin{quote}
  ``... recursion can be encoded using function storage, as noted by Landin (folklore).'' ~\cite{levy2002}
\end{quote}
\begin{quote}
  ``... in languages with higher-order store, it is usually possible to write recursion operators by backpatching function pointers.'' ~\cite{krishnaswami2015}
\end{quote}
\noindent 
However, the actual cause of the recursion is due to
the update to the \emph{function’s environment} through the store, not
by storing the function in a reference.
Even more interestingly, this update to the function’s environment actually requires
\textit{impredicativity} to be well typed, but this impredicativity
is implicit and hidden in the usual function type.
This impredicativity exists in function types even if not explicit in the type system.
We conjecture that restricting this impredicativity is one way to recover strong
normalization, without the use of linear types.

We can make this observation more precise by closure converting the example.
Closure conversion transforms functions into explicit closures, that is, pairs
of closed procedure code with its environment.
To type closures, we use the usual typing from \citet{minamide1996}, where closures
are typed as existential pairs, abstracting the type of the environment.
The environment is a tuple of the free variables from the procedure code.

\vspace{-2ex}

\begin{displaymath}
  \begin{array}[t]{l}
    id : \<exists> \alpha . \langle (\<Nat> \to \alpha \to \<Nat>) \times \alpha \rangle
    \\
    id = \<pack> \langle \langle \rangle, \langle (\<lam> x .\<lam> env . x), \langle \rangle \rangle \rangle
    \\
    r : \<Ref> (\<exists> \alpha . \langle (\<Nat> \to \alpha \to \<Nat>) \times
    \alpha \rangle) \qquad \texttt{-- abbreviated \<Reft> below}
    \\
    r = \<new> id
    \\
    f : \<exists> \alpha . \langle (\<Nat> \to \alpha \to \<Nat>) \times \alpha \rangle
    \\
    f = \<pack> \langle \langle \<Reft> \rangle , \langle (\<lam> x . \<lam> env : \langle \<Reft> \rangle. \<let> r = \<proj1> env \<in> (\<deref> r) \ x ), \langle r \rangle \rangle \rangle
    \\ r := f ; (\<proj1> f) \ 0 \ (\<proj2> f)
  \end{array}
\end{displaymath}

\vspace{-1ex}

The functions $f$ and $id$ have the same type as closures, since the type of the
environment is abstract.
The $r$ reference is then updated with a closure that contains $r$ in its
environment, and is well typed precisely because of the
(implicit) impredicativity of the existential pair.
The sort of universe of $\alpha$ is unrestricted, as is normal in System F,
meaning the existentially quantification variable $\alpha$ can include the existential
itself, or a reference that contains the existential.




\section{Background and Related work}
Related work modelling references often divide references into three categories.
\textit{Ground} references store only base types, \textit{full-ground}
\cite{murawski2018} (or sometimes called recursive) references store base types
and other references, and higher-order references store unrestricted
higher-order functions.
With the closure-converted Landin’s Knot example, we see a middle ground between a
terminating language with a full-ground store and a non-terminating language with
a higher-order store.

The impredicativity requirement to type Landin’s Knot has been recognized,
but unexplored, in past work modeling languages with higher-order references.
\citet{levy2002} describes a possible world semantics to model languages with these
three different kinds of references, and modeling functions in the store requires
recursive domain equations.
\citet{kammar2017} model full-ground references, but observe that in order to
model higher-order references, recursive domain equations or step indexing is required.
\citet{ahmedthesis} notes the circularity of modeling a store as mapping a location
to a type, and that types are modelled as a predicates on stores, explicitly
noting that the impredicative quantification of existential types is related to
this circularity, but uses step-indexing to side-step this circularity.
All this prior work describe possible models for higher-order references, but
they do not investigate the alternative to impredicative quantification over a
function's environment.
Thus, our proposal is: design a language with higher-order references that is
terminating by requiring predicativity with respect to environments and by
interpreting the store inductively.
The store at one level will depend only on the interpretation of terms at a lower
universe, with a ground store as the base case.

%% file: proposal.tex
\section{Proposal for language design}
\label{sec:proposal}

The closure-converted Landin's Knot shows us that the non-termination is due to
updates to the function's environment.
This guides us in our language design proposal; there needs to be some restriction
on the types of the free variables a function closes over.
We use the $::$ annotation to indicate the sort of a type.

To see how we might restrict the environment, consider the (perhaps overly)
restrictive approach where we only allow full-ground references to appear in the
environment.
Below are a possible typing rule for closures and the (back-translated) equivalent
rule for source language functions.
\begin{mathpar}
  \inferrule
  {\Gamma \vdash \tau :: \<fg> \\
   \Gamma \vdash e : \subst{\tau_1}{\tau}{\alpha}}
  {\Gamma \vdash \<pack> \langle \tau, e \rangle : \<exists> \alpha . \tau_1}

  \inferrule
  {\Gamma \vdash FV(e) : \tau :: \<fg> \cdots \\
   \Gamma, x : \tau_1 \vdash e : \tau_2}
  {\Gamma \vdash \<lam> x : \tau_1 . e : \tau_1 \to \tau_2}
\end{mathpar}
In both rules, we use the sort to restrict the type of the environment.
Either the type of the environment $\tau$ is restricted, or the types of free
variables in the body of a function are restricted to be of a full-ground sort.
Since closures are not ground types, Landin's Knot cannot be well typed.
References can store closures, but the closures \textit{themselves}
are restricted.
In a source language with references, this restricts the free variables in the
function body to be full-ground.
The downside of this approach is that higher-order closures would not be possible to
encode, assuming that all closures must be allocated in mutable cells (heap allocated).

An alternative approach is to restrict the quantification over
environments in closures.
Each environment is typed at a the highest universe level of the types
in the environment.
\begin{mathpar}
  \inferrule
  {\Gamma \vdash \tau :: \<Type>_j \\
   \Gamma \vdash e : \subst{\tau_1}{\tau}{\alpha}}
  {\Gamma \vdash \<pack> \langle \tau, e \rangle : \<exists> \alpha : \<Type>_j . \tau_1}

  \inferrule
  {\Gamma, \alpha : \<Type>_j \vdash \tau_1 :: \<Type>_k}
  {\Gamma \vdash \<exists> \alpha : \<Type>_j . \tau_1 :: \<Type>_k}
  
  \inferrule
  {\Gamma \vdash FV(e) : \tau :: \<Type>_i \cdots \\
   \<max>(\hdots i \hdots) = j  \\   
   \Gamma, x : \tau_1 \vdash e : \tau_2 :: \<Type>_k}
  {\Gamma \vdash \<lam> x : \tau_1 . e : \tau_1 \to \tau_2 :: \<Type>_j}
\end{mathpar}
Existential types in the target language are standard---impredicative, which is
necessary to type higher-order closures, but with an explicit sort annotation on
the existentially bound variable.
Note that for closures in particular, because the environment shows up in type
$\tau_1 = C \times \alpha$, a closure with environment $\alpha : \<Type>_j$
must live in $\<Type>_j$ as well.
The means closures can capture other closures, including closures of the same
type, as is expected by closure conversion.

Functions are impredicative in their parameter, but predicative in the types of
their environment variables; we've combined the typing and sort of functions
above to express this.
This makes the sort of the environment explicit in the source language typing
rule, and forces a function's sort to be the same as its environment, as in
the closure converted language.

Finally, we add the following sort rule for references, which breaks the
circularity when a reference contains a closure, preventing an environment from
containing a reference that contains its own type.
\begin{mathpar}
  \inferrule
  {\Gamma \vdash A :: \<Type>_i}
  {\Gamma \vdash \<Ref> A :: \<Type>_{i+1}}
\end{mathpar}
\noindent
Base types are at level $0$, but references bump up the level of
the stored type by $1$.
A reference cannot be updated to contain a closure that contains itself, since
the type level of the environments would be necessarily lower than the type of
the reference that must contain the environment.
A closure can capture its own type in its environment, but it \emph{cannot}
capture a \<Ref> of its own type.\footnote{We give a full derivation
in~\autoref{appendix}.}
Typing our earlier example with these rules, the environment of the closure $f$
contains a reference, so its environment type is forced to be of sort
$\<Type>_1$, while $id$'s environment type is $\<Type>_0$.
This approach is similar to the step-indexing model developed by \citet{ahmedthesis},
but makes type levels explicit in the type system, rather than in the meta-theory.

We have yet to prove any of these proposed languages terminating, and are still
open to exploring other designs as well.
But the intuition is that the semantics of stores will be inductive on the
universe level, with each level able to contain only functions that close over
the previous level of stores.
At level 0 the semantics is equivalent to a
ground store, and so the store must contain only terminating functions.
Our eventual goal is to extend this language design not only to a simply typed
language, but also to a dependently typed language like the Calculus of Constructions (CC).
This could serve as an IL for a type-preserving compiler from CC to a low-level
language like C, enabling a compiler with a type-check-then-link approach
\cite{bowmanthesis}.

%% file: appendix.tex
\begin{landscape}
\section{Appendix}
\label{appendix}

Below we have the tree for deriving that $id$ has type $\<exists> \alpha : \<Type>_0. (\<Nat> \to \alpha \to \<Nat>) \times \alpha$, using our proposed typing rules.

\begin{displaymath}
\inference
    {\Gamma \vdash \langle \rangle :: \<Type>_0
      &
      \inference
          {\Gamma \vdash  (\<lam> x .\<lam> env . x) : \<Nat> \to \langle \rangle \to \<Nat> :: \<Type>_0 &
           \Gamma \vdash \langle \rangle : \langle \rangle :: \<Type>_0}
                {\Gamma \vdash  \langle (\<lam> x .\<lam> env . x), \langle \rangle \rangle : \subst{(\<Nat> \to \alpha \to \<Nat>) \times \alpha}{\langle \rangle}{\alpha}}}
    {\Gamma \vdash \<pack> \langle \langle \rangle, \langle (\<lam> x .\<lam> env . x), \langle \rangle \rangle \rangle : \<exists> \alpha : \<Type>_0. (\<Nat> \to \alpha \to \<Nat>) \times \alpha :: \<Type>_0}
\end{displaymath}

Below we have the tree for deriving that $f$ has type $\<exists> \alpha : \<Type>_1. (\<Nat> \to \alpha \to \<Nat>) \times \alpha$, using our proposed typing rules.
The function body has been abbreviated to $e$.
$\<Reft>$ is $\<Ref> (\<exists> \alpha : \<Type>_0. (\<Nat> \to \alpha \to \<Nat>) \times \alpha)$, which we expand when we derive its sort, but abbreviate otherwise.
We omit the subtree of deriving the type for the function body $e$, as it is similar to the $id$ case and not interesting.
The proposed rule for $\<Ref>$ bumps the sort level of the environment of $f$, which
in turn bumps the sort level for the pair of code and environment, since the
environment contains a reference.

\begin{displaymath}
\inference
    {\inference
      {\texttt{by derivation $D_1$}}
      {\Gamma \vdash \langle \<Reft> \rangle :: \<Type>_1} &
      \inference
          {\Gamma \vdash  (\<lam> x . \<lam> env : \langle \<Reft> \rangle. e ) : \<Nat> \to \langle \<Reft> \rangle \to \<Nat> :: \<Type>_1 &
            \inference
                {\texttt{by derivation $D_1$}}
                {\Gamma \vdash \langle \<Reft> \rangle :: \<Type>_1} }
          {\Gamma \vdash  \langle  (\<lam> x . \<lam> env : \langle \<Reft> \rangle. e ), \langle r \rangle \rangle : \subst{(\<Nat> \to \alpha \to \<Nat>) \times \alpha}{\langle \<Reft> \rangle}{\alpha} :: \<Type>_1}}
    {\Gamma \vdash \<pack> \langle \langle \<Reft> \rangle , \langle (\<lam> x . \<lam> env : \langle \<Reft> \rangle. e ), \langle r \rangle \rangle \rangle : \<exists> \alpha : \<Type>_1. \langle (\<Nat> \to \alpha \to \<Nat>) \times \alpha \rangle :: \<Type>_1}
\end{displaymath}
\noindent where $D_1$ is
\begin{displaymath}
    \inference{
     \inference[Ref]
        {\inferrule
          {\Gamma \vdash \<Type>_0 :: \<Type>_1 \and
            \Gamma, \alpha :: \tikzmark{a}{\colorbox{lightgray}{$\<Type>_0$}} \vdash (\<Nat> \to \alpha \to \<Nat>) \times \alpha :: \tikzmark{b}{\colorbox{lightgray}{$\<Type>_0$}}}
          {\Gamma \vdash \<exists> \alpha : \<Type>_0. (\<Nat> \to \alpha \to \<Nat>) \times \alpha :: \tikzmark{c}{\colorbox{lightgray}{$\<Type>_0$}}}}
        {\Gamma \vdash \<Ref> (\<exists> \alpha : \<Type>_0. (\<Nat> \to \alpha \to \<Nat>) \times \alpha) :: \tikzmark{d}{\colorbox{lightgray}{$\<Type>_1$}}}}
              {\Gamma \vdash \langle \<Reft> \rangle :: \<Type>_1}
\end{displaymath}

Because of the differing sort annotations, the update to the reference $r$ cannot
be well typed.
This prevents the non-termination caused by backpatching.

More generally, the derivation of $D_1$ shows how the type of the \<Ref> must
be one greater than the environment of the closure it contains, so the \<Ref>
cannot appear in the environment of a closure it contains.
\end{landscape}